\documentclass[conference]{IEEEtran}

\usepackage{array}
\usepackage[caption=false,font=normalsize,labelfont=sf,textfont=sf]{subfig}
\usepackage{textcomp}
\usepackage{stfloats}
\usepackage{url}
\usepackage{verbatim}
\usepackage{graphicx}
 
\usepackage{cite}
\usepackage{amsmath,graphicx}
\usepackage{amssymb}
\usepackage{algpseudocode}
\usepackage{amsfonts}
\usepackage{fancyhdr}  %
\usepackage{cases}
\usepackage{extarrows}
\usepackage{algorithm}
\usepackage{multirow,tabularx}
\usepackage{mathtools}
\usepackage{xcolor}
\usepackage[english]{babel}
\usepackage{caption}
\usepackage{autobreak}
\usepackage{bbm}

\ifCLASSINFOpdf
\else
\fi
\begin{document} 
\title{Multi-User Wireless Communications with Holographic MIMO Surfaces: A Convenient Channel Model and Spectral Efficiency Analysis}

	\author{
	\IEEEauthorblockN{Li Wei\IEEEauthorrefmark{1}, Chongwen Huang\IEEEauthorrefmark{2},George~C.~Alexandropoulos\IEEEauthorrefmark{3}, \\Wei E. I. Sha\IEEEauthorrefmark{2}, Zhaoyang Zhang\IEEEauthorrefmark{2}, M\'{e}rouane~Debbah\IEEEauthorrefmark{4} and Chau~Yuen\IEEEauthorrefmark{1} }
	\IEEEauthorblockA{\IEEEauthorrefmark{1}Singapore University of Technology and Design, 487372 Singapore}
	\IEEEauthorblockA{\IEEEauthorrefmark{2}College of Information Science and Electronic Engineering, Zhejiang University }
	\IEEEauthorblockA{\IEEEauthorrefmark{3}Department of Informatics and Telecommunications, National and Kapodistrian University of Athens, Greece}
	\IEEEauthorblockA{\IEEEauthorrefmark{4}Technology Innovation Institute, Masdar City, Abu Dhabi, United Arab Emirates}
	
}

\maketitle

\begin{abstract}
The multi-user Holographic Multiple-Input and Multiple-Output Surface (MU-HMIMOS) paradigm, which is capable of realizing large continuous apertures with minimal power consumption and of shaping radio wave propagation at will, has been recently considered as an energy-efficient solution for future wireless networks. The tractable channel modeling of MU-HMIMOS signal propagation is one of the most critical challenges, mainly due to the coupling effect induced by the excessively large number of closely spaced patch antennas. In this paper, we focus on this challenge for downlink communications and model the electromagnetic channel in the wavenumber domain using the Fourier plane wave representation. Based on the proposed model, we devise a Zero-Forcing (ZF) precoding scheme, capitalizing on the sampled channel variance that depends on the number and spacing of the HMIMOS patch antennas, and perform a spectral efficiency analysis. Our simulation results showcase that the more patch antennas and the larger their spacing is, the performance of the considered MU-HMIMOS system improves.  In addition, it is demonstrated that our theoretical performance expressions approximate sufficiently well the simulated spectral efficiency, even for the highly correlated cases, thus verifying the effectiveness and robustness of the presented analytical framework. 
\end{abstract}


\section{Introduction}\label{sec:intro}
In recent years, the demand for ubiquitous wireless communications is growing due to the explosive development of mobile devices and multimedia applications \cite{9374451}. To satisfy the expanding demands, the exploitation of the ultimate limits of communication techniques is concerned, and new technologies to provide higher bandwidth and energy efficiency also attract great interest, e.g., TeraHertz (THz) \cite{9325920} and millimeter-Wave (mmWave) communications \cite{Akyildiz2018mag}, as well as extreme Multiple-Input Multiple-Output (MIMO) catering for massive users and throughput \cite{shlezinger2020dynamic_all,9475156}. However, more profound requirements for data rate and massive connections are needed for 6-th Generation (6G) wireless communications. In addition, the exploitation of higher frequencies and larger bandwidths brings many challenges. 

Fortunately, the development of Reconfigurable Intelligent Surface (RIS) and metasurfaces provides a feasible and engaging research direction towards realizing highly flexible antennas at low cost  \cite{yanglianghnu20, 9501003,8972400,9475160,9366805}.  Specifically, an RIS is usually comprised of a large number of hardware-efficient and nearly passive reflecting elements, each of which can alter the phase of the incoming signal without requiring a dedicated power amplifier \cite{holobeamforming,huang2019reconfigurable,WavePropTCCN}. Inspired by the potential of RISs in 6G communications, the Holographic MIMO Surface (HMIMOS) concept aims at going beyond massive MIMO \cite{9136592, demir2021channel}. Typically, HMIMOS incorporates densely packed sub-wavelength patch antennas to achieve programmable wireless environments \cite{RISE6G_COMMAG}, which is verified to boost competitiveness in many fields, including NOMA, unmanned aerial vehicles, mmWaves, and multi-antenna systems \cite{9530717}. \cite{9136592} proved the flexibility of HMIMOS configuration and the advantages in improving SE through intelligent environment configuration. Authors in \cite{9145091} investigated the mutual coupling matrix in large RIS-assisted single user communication system, and showed that large RIS could achieve super-directivity. Benefiting from these merits, HMIMOS can be integrated into many fields, including the extension of coverage, wireless power transfer, and indoor positioning \cite{9136592,alexandg_2021}. However, the full exploitation of HMIMOS is still a challenge due to many non-trivial issues.  

One of the main challenges is multi-user channel modeling due to the spatially continuous aperture realized by holographic patch antennas \cite{9475156}. The traditional channel models are not applicable for various reasons, e.g., the coupling between antennas, their massive numbers, and the surface-to-surface transmission. Thus, the shift towards infinite antennas and operating frequency beyond the THz requires an effective EM model, since traditional independent Rayleigh fading models based on the assumption of far-field electromagnetic propagation might be not applicable \cite{8437634, sanguinetti2021wavenumber,PhysFad}. In this regime, the spatially correlated random field cannot be ignored, and a tractable stochastic tool is required to model the channel. In \cite{9110848,arxivjstsp}, a spatially correlated small-scale fading model for single user is proposed employing Fourier plane wave representation to generate samples of the random field computationally. Meanwhile, it is also challenging to efficiently analyze the spatially-continuous EM channel theoretically. The work in \cite{9154219} investigated the number of channel spatial Degrees of Freedom (DoF) in isotropic scattering environments considering spatially-constrained apertures, and it was proved that the spatial DoF is proportional to the surface area, which is distinct from traditional analysis methods. 
 
Motivated by EM channel modeling for single-user in \cite{9110848}, this paper investigates channel modeling for MU-HMIMOS communication systems, and theoretically analyzes the SE of various linear precoding schemes. Typically, the wireless channel empowered by an HMIMOS with infinite patch antennas is assumed to be on continuous EM space in the optimal setting. To facilitate practical applications, the continuous EM channel is sampled according to a specific placement of the array patch antennas. Thus, the communication is considered as a functional analysis problem depends only on geometric relationships, and the involved channel is constructed as a combination of complete basis function set of transmitter/receiver surfaces. In this manner, the ultimate performance for communications, namely the intrinsic capacity of the sampled space wireless channel, can be studied. Specifically, due to the EM channel being characteristic of complete basis functions, the capacity can be theoretically analyzed using the variances of the sampled transmitter/receiver surfaces. 

The remainder of this paper is organized as follows. In Section \ref{sec:EM_channel}, the proposed EM channel modeling for MU-HMIMOS communication systems is presented.  Section~\ref{sec:sum_rate} presents the analytical formulas for the SE with the ZF precoding scheme. Our simulation results are given in Section~\ref{sec:simu}, while the concluded remarks of the paper are drawn in Section~\ref{sec:conclusion}.
	
\textit{Notation}: Fonts $a$, $\mathbf{a}$, and $\mathbf{A}$ represent scalars, vectors, and matrices, respectively. $\mathbf{A}^T$, $\mathbf{A}^H$, $\mathbf{A}^{-1}$, $\mathbf{A^\dag}$, and $\|\mathbf{A}\|_F$ denote transpose, Hermitian (conjugate transpose), inverse, pseudo-inverse, and Frobenius norm of $ \mathbf{A} $, respectively.  $ \mathbf{A}_{i,j}$ or $[\mathbf{A}]_{i,j}$ represents $\mathbf{A}$'s $(i,j)$-th element, while $[\mathbf{A}]_{i,:}$ and $[\mathbf{A}]_{:,j}$ stand for its $i$-th row and $j$-th column, respectively. $|\cdot|$ and $(\cdot)^*$ denote the modulus and conjugate, respectively. $\text{tr}(\cdot)$ gives the trace of a matrix, $\mathbf{I}_n$ (with $n\geq2$) is the $n\times n$ identity matrix, and $\mathbf{1}_n$ is a column vector with all ones.   $\delta_{k,i}$ equals to $1$ when $k=i$ or $0$ when $k\neq i$, and $\mathbf{e}_n$ is the $n$-th unit coordinate vector with $1$ in the $n$-th basis and $0$'s in each $n^\prime$-th basis $\forall$$n^\prime \neq n$. Finally, notation ${\rm diag}(\mathbf{a})$ represents a diagonal matrix with the entries of $\mathbf{a}$ on its main diagonal,  $\delta(\cdot)$ is the Dirac delta function, and $\odot$ is the Hadamard product.

\section{ EM Channel Modeling for MU-HMIMOS} \label{sec:EM_channel}
In this section, an EM-compliant far-field channel model for multi-user HMIMOS communication systems using approximated Fourier plane-wave series expansion is presented. 
\subsection{System Model}\label{subsec:signal model}
\begin{figure} 
	\begin{center}
		\centerline{\includegraphics[width=0.4\textwidth]{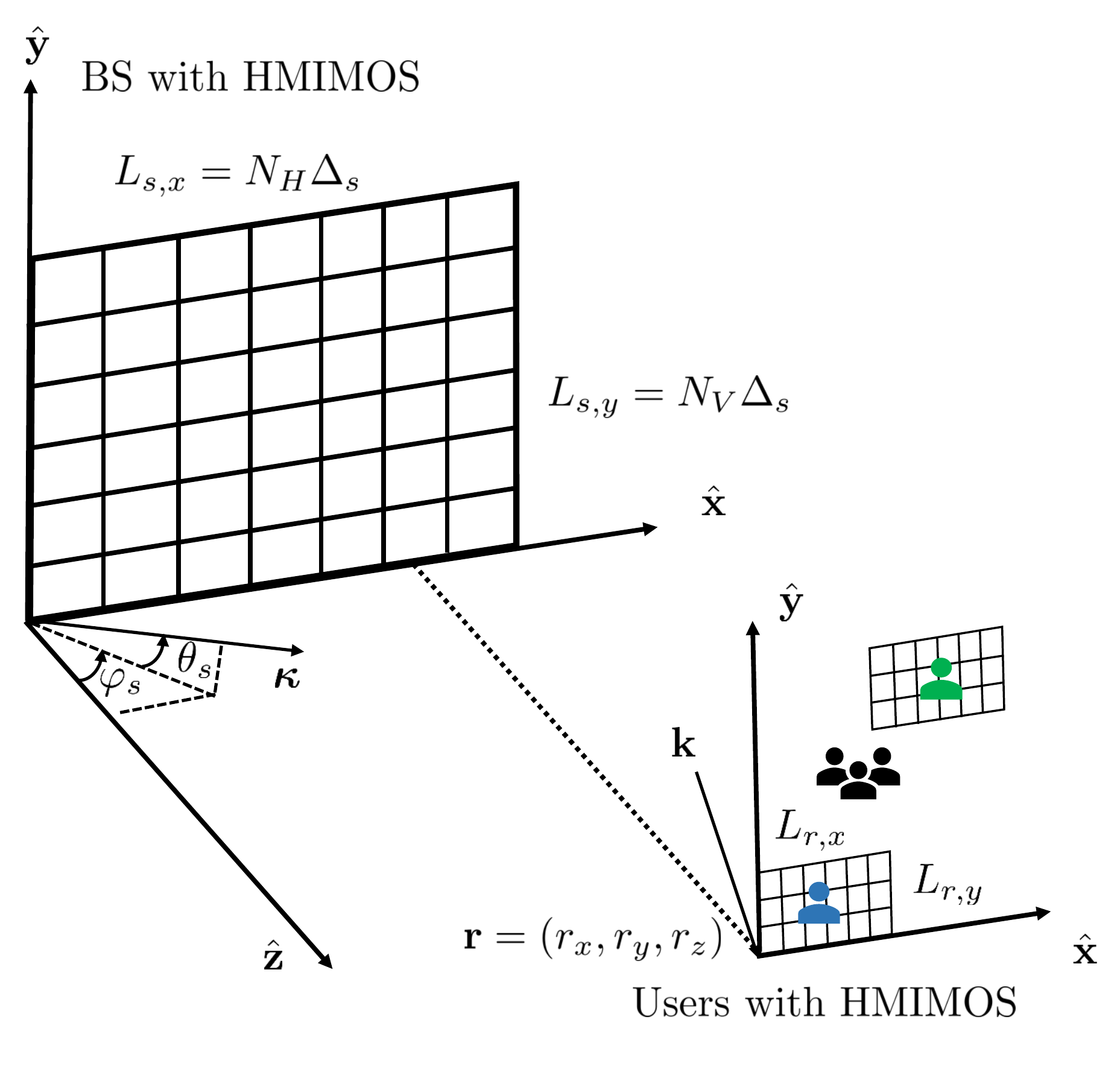}}  \vspace{-0mm}
		\caption{The considered MU-HMIMOS communication system consisting of an $N_s$ patch antennas at BS end and $M$ users with each having $N_r$ patch antennas.}
		\label{fig:Estimation_Scheme} \vspace{-8mm}
	\end{center}
\end{figure}

Consider the downlink communication between a BS and a group of $M$ users that are both equipped with HMIMOS,  as shown in Fig$.$~\ref{fig:Estimation_Scheme}. The HMIMOS at BS side is comprised of $N_s=N_V N_H$  unit cells, each made from metamaterials that are capable of adjusting their reflection coefficients, and the patch antennas spacing $\Delta_s$ is below half of the wavelength $\lambda$. The horizontal and vertical length of HMIMOS is $L_{s,x}= N_H \Delta$ and $L_{s,y}=N_H \Delta_s$. Each user is equipped with $N_r$ patch antennas with spacing $\Delta_r$, and $L_{r,x}$ and $L_{r,y}$ are the horizontal and vertical length, respectively.  

\subsection{Channel Modeling for Individual User}
There exists transformation between the space domain channel $\mathbf{H}$ and wavenumber domain channel $\mathbf{H}_a$ \cite{9110848}.  Specifically,  the $(m,n)$-th entry of channel matrix $\mathbf{H} \in \mathbb{C}^{N_r \times N_s}$ is  \cite{9110848}
\begin{equation}
	\begin{aligned}
		[\mathbf{H}]_{mn}&\!=\!\frac{1}{(2 \pi)^{2}} \iiiint_{\mathcal{D} \times \mathcal{D}} \! {a}_{r,m}(\mathbf{k}, \mathbf{r}) H_{a}\!\left(k_{x}, k_{y}, \kappa_{x}, \kappa_{y}\right) \\
		&\qquad  \qquad \qquad {a}_{s,n}(\boldsymbol{\kappa}, \mathbf{s}) d k_{x} d k_{y} d \kappa_{x} d \kappa_{y},
	\end{aligned}
\end{equation} 
where ${a}_{s,n}(\boldsymbol{\kappa}, \mathbf{s})$ is the $n$-th element in  transmit vector, and ${a}_{r,m}(\mathbf{k}, \mathbf{r})$ is the $m$-th element in receive vector, and integration region is $\mathcal{D}=\{(k_x,k_y) \in \mathbb{R}^2: k_x^2+k_y^2 \leq \kappa ^2 \}$. The equivalent wavenumber domain channel is
\begin{equation}
	\begin{aligned}
		&H_{a}\!\left(\!k_{x}, k_{y}, \!\kappa_{x},\! \kappa_{y}\!\right)\!=\!S^{1/2}\!\left(k_{x},\! k_{y},\! \kappa_{x},\! \kappa_{y}\right) \!W\!\left(k_{x},\! k_{y}, \! \kappa_{x}, \! \kappa_{y}\right)\\
		&=\frac{\kappa \eta}{2} \frac{  A\left(k_{x}, k_{y}, \kappa_{x}, \kappa_{y}\right) W\left(k_{x}, k_{y}, \kappa_{x}, \kappa_{y}\right)}{k_{z}^{1/2}\left(k_{x}, k_{y}\right) \kappa_{z}^{1/2} \left(\kappa_{x}, \kappa_{y}\right)},
	\end{aligned}
\end{equation}
where the spectral factor $A\left(k_{x}, k_{y}, \kappa_{x}, \kappa_{y}\right)$ is an arbitrary real-valued, non-negative function, and $W\left(k_{x}, k_{y}, \kappa_{x}, \kappa_{y}\right)$ is a collection of unit-variance independent and i.i.d. circularly-symmetric and complex-Gaussian random variables. i.e., $W\left(k_{x}, k_{y}, \kappa_{x}, \kappa_{y}\right)\sim \mathcal{CN}(0,1)$.

The Fourier plane-wave series expansion is non-zero only within the lattice ellipse
\begin{equation}
	\begin{aligned}
		&\mathcal{E}_{s}\!=\!\left\{\!\left(m_{x}, m_{y}\right) \in \mathbb{Z}^{2}\!:\!\left(\!m_{x} \lambda / L_{s,x}\!\right)^{2}\!+\!\left(m_{y} \lambda / L_{s,y}\right)^{2} \!\leq\! 1\!\right\}\!, \\
		&\mathcal{E}_{r}\!=\!\left\{\left(\ell_{x}, \ell_{y}\right) \in \mathbb{Z}^{2}\!:\left(\ell_{x} \lambda / L_{r, x}\right)^{2}+\left(\ell_{y} \lambda / L_{r, y}\right)^{2} \!\leq 1\!\right\},
	\end{aligned}
\end{equation}
at the source and receiver, respectively.  
Adopting Fourier plane-wave series expansion, we have
\begin{equation}
\begin{aligned}
	[\mathbf{H}]_{mn} \approx &\sum_{\left(\ell_{x}, \ell_{y}\right) \in \mathcal{E}_{r}} \sum_{\left(m_{x}, m_{y}\right) \in \mathcal{E}_{s}}  H_{a}\left(\ell_{x}, \ell_{y}, m_{x}, m_{y}\right)\\
	&\qquad  {a_{r,m}\left(\ell_{x}, \ell_{y}, \mathbf{r}\right) a_{s,n}\left(m_{x}, m_{y}, \mathbf{s}\right)},
\end{aligned}	 
\end{equation}
where the Fourier coefficient at the sampling point at $(m_x,m_y,\ell_{x},\ell_{y})$ is  
\begin{equation}
	H_{a}\left(\ell_{x}, \ell_{y}, m_{x}, m_{y}\right) \sim \mathcal{N}_{\mathbb{C}}\left(0, \sigma^{2}\left(\ell_{x}, \ell_{y}, m_{x}, m_{y}\right)\right),
\end{equation}
with the variance given by 
\begin{equation} \label{equ:S_int}
\begin{aligned}
	\frac{1}{(2 \pi)^{4}}  \iiiint_{\mathcal{S}_{s}  \times \mathcal{S}_{r} } S\left(k_{x}, k_{y}, \kappa_{x}, \kappa_{y}\right)  d k_{x} d k_{y} d \kappa_{x} d \kappa_{y},
\end{aligned}
\end{equation}
where the sets $\mathcal{S}_{s}$ and $\mathcal{S}_{r}$ are defined as 
\begin{equation}
	\begin{aligned}
		&\left\{\left[\frac{2 \pi m_{x}}{L_{s, x}}, \frac{2 \pi\left(m_{x}+1\right)}{L_{s, x}}\right] \times\left[\frac{2 \pi m_{y}}{L_{s, y}}, \frac{2 \pi\left(m_{y}+1\right)}{L_{s, y}}\right]\right\}, \\
		&\left\{\left[\frac{2 \pi \ell_{x}}{L_{r, x}}, \frac{2 \pi\left(\ell_{x}+1\right)}{L_{r, x}}\right] \times\left[\frac{2 \pi \ell_{y}}{L_{r, y}}, \frac{2 \pi\left(\ell_{y}+1\right)}{L_{r, y}}\right]\right\}.
	\end{aligned}
\end{equation} 
  
\subsection{Channel Modeling for Multiple Users}\label{sec:iso_channel}
This part extends the individual user channel modeling from the previous subsection to the multi-user case, we assume that different users are independently distributed in space, thus, the multi-user involved channel matrix can be decomposed into multiple channel matrices that corresponds to different users. For simplicity, the isotropic scattering environment is considered. The channel of $m$-th user in matrix form $\mathbf{H}^{(m)} \in \mathbb{C}^{N_r \times N_s}$ is 
\begin{equation}
\begin{aligned}
	&\mathbf{H}^{(m)} \! = \!\sqrt{N_r \! N_s} \! \sum_{\left(\!\ell_{x}, \ell_{y}\!\right) \!\in \mathcal{E}_{\!r}} \sum_{\left(\!m_{x}, m_{y}\!\right) \!\in \!\mathcal{E}_{\!s}} H_{a}^{(\!m\!)}\left(\ell_{x},\! \ell_{y},\! m_{x},\! m_{y}\right) \\
	& \qquad\qquad {\mathbf{a}_{r}\left(\ell_{x}, \ell_{y}, \mathbf{r}^{(m)}\right) \mathbf{a}_{s}^{H}\left(m_{x}, m_{y}, \mathbf{s}\right)} ,
\end{aligned}	
\end{equation}
where the Fourier coefficient 
\begin{equation}
	H_{a}^{(m)}\left(\ell_{x}, \!\ell_{y},\! m_{x},\! m_{y}\right) \!\sim\! \mathcal{N}_{\mathbb{C}}\left(0, \sigma^{2}_{(m)}\left(\ell_{x}, \!\ell_{y}, \!m_{x},\! m_{y}\right)\right),
\end{equation}
and the sampling points at $(m_x,m_y)$ and $(\ell_{x},\ell_{y})$ are 
\begin{equation}
	\begin{aligned}
		&[a_{s}\left(m_{x},\! m_{y}, \!\mathbf{s}\right)]_j\!= \!\frac{1}{\!\sqrt{N_s}\!} e^{\!-\!\mathrm{j}\!\left(\!\frac{2 \!\pi\!}{L_{\!s\!,\!x}} \!m_{x} \!s_{\!x_j}\!+\!\frac{2 \!\pi}{L_{\!s\!,\!y}} \!m_{\!y}\! s_{\!y_j\!}\!+\!\gamma_{\!s}\!\left(\!m_{\!x}\!, \!m_{\!y}\right)\! s_{z_j}\!\right)\!}\!,\\
		&\qquad \qquad \qquad  j=1,\ldots, N_s, \\
		&[a_{r}\!\left(\ell_{x}\!, \!\ell_{y}, \!\mathbf{r}^{(m)}\!\right)\!]_i\!=\! \frac{1}{\!\sqrt{N_r}\!} e^{\!\mathrm{j}\!\left(\!\frac{2 \!\pi}{L_{r,\! x}} \!\ell_{x}\! r_{x_i}^{(m)}\!+\!\frac{2 \!\pi}{\!L_{r, \!y}} \!\ell_{y}\! r_{y_i}^{\!(m)\!}\!+\!\gamma_{\!r}\!\left(\!\ell_{x},\! \ell_{y}\!\right) \!r_{z_i}^{(\!m\!)}\right)\!}\!,\\
		& \qquad \qquad \qquad  i=1,\ldots, N_r.
	\end{aligned}
\end{equation}

The $N_s$ entries $[a_{s}\left(m_{x}, m_{y}, \mathbf{s}\right)]_j$ are collected in $\mathbf{u}_s(m_x,m_y)$, and $N_r$ entries $[a_{r}\left(\ell_{x}, \ell_{y}, \mathbf{r}^{(m)}\right)]_i$ are collected in $\mathbf{u}_r^{(m)}(\ell_{x}, \ell_{y} )$. Denote $\mathbf{U}_s \in \mathbb{C}^{N_s \times n_s}$ as the matrix collecting the $n_s$ column vectors $\mathbf{u}_s(m_x,m_y)$, and  $\mathbf{U}_r^{(m)} \in \mathbb{C}^{N_r \times n_r}$ as the matrix collecting the $n_r$ column vectors $\mathbf{u}_r^{(m)}(\ell_{x}, \ell_{y})$. Specifically, $\mathbf{U}_s  \mathbf{U}_s^{H}=\mathbf{I}_{n_s}$ and $\mathbf{U}_r^{(m)}  \mathbf{U}_r^{{(m)} H}=\mathbf{I}_{n_r}$ since their columns describe discretized transmit and receive plane-wave harmonics.

Based on the above observation, the channel matrix can be approximated by spatial sampling $n_s$ and $n_r$ points at $j=1,\ldots,N_s$-th patch antenna and the $i=1,\ldots,N_r$-th received antenna for $m$-th user. Thus,
\begin{equation}
	\begin{aligned}
		\mathbf{H}^{(m)}=\mathbf{U}_r^{(m)} \mathbf{H}_a^{(m)}  \mathbf{U}_s^{H}=\mathbf{U}_r^{(m)} \left(\mathbf{\Sigma}^{(m)} \odot \mathbf{W}\right)  \mathbf{U}_s^{H},
	\end{aligned}
\end{equation}
where $\mathbf{H}_a^{(m)} = \mathbf{\Sigma}^{(m)} \odot \mathbf{W} \in \mathbb{C}^{n_r\times n_s}$ collects all $\sqrt{N_r N_s} 	H_{a}^{(m)}\left(\ell_{x}, \ell_{y}, m_{x}, m_{y}\right)$, with $\mathbf{\Sigma}^{(m)}$ collects all entries $N_r N_s \sigma^{2}_{(m)}\left(\ell_{x}, \ell_{y}, m_{x}, m_{y}\right)$  and $\mathbf{W}\sim \mathcal{CN}(0,\mathbf{I}_{n_r n_s})$.

The above derivation focuses on the $m$-th user channel matrix, here, we consider $M$ users in a general form, we assume that distances between different users are large enough to dependently model channels among users. 

Thus, the channel $\mathbf{H}\in\mathbb{C}^{N_r M \times N_s}$ is given by 
\begin{equation}
	\begin{aligned}
		\mathbf{H}& = \left[ \mathbf{U}_s^{*} {\mathbf{H}_a^{(1)}}^{T} {\mathbf{U}_r^{(1)}}^{T} ,\ldots,\mathbf{U}_s^{*} {\mathbf{H}_a^{(M)}}^{T} {\mathbf{U}_r^{(M)}}^{T}  \right]^{T} \\
		&=\mathbf{U}_r \mathbf{H}_a   \mathbf{U}_s^{H}.
	\end{aligned}
\end{equation} 
 
It can be observed that the complex spatially correlated channel matrix $\mathbf{H}^{(m)}\in \mathbb{C}^{N_r \times N_s}$ is equivalent to a low-dimensional wavedomain channel matrix $\mathbf{H}_a^{(m)}\in \mathbb{C}^{n_r \times n_s}$.

\section{Capacity evaluation} \label{sec:sum_rate}
In this section, we derive the SE of the downlink MU-HMIMOS communication system adopting ZF precoding scheme.  Different from the conventional precoding analysis of MIMO system, where each element in channel matrix is normally assumed to have the same unit variance, in this work, each entry in the constructed EM channel has different variances. Thus, the traditional analysis methods  cannot be directly applied in the considered scenario.

The HMIMOS at the transmitter is composed of $N_s$ elements, and each user is equipped with $N_r$ patch antennas. There are total $M$ users, thus, the downlink MU-HMIMOS wireless communication system model is
\begin{equation}
\begin{aligned}
	\mathbf{y}&=\sqrt{p_u} \mathbf{H}   \mathbf{\Phi} \mathbf{V} \mathbf{x}+\mathbf{w}=\sqrt{p_u} \mathbf{U}_r  \mathbf{H}_a \mathbf{U}_s^{H} \mathbf{\Phi} \mathbf{V}  \mathbf{x} +\mathbf{w},
\end{aligned}
\end{equation}
where $\mathbf{y}\in \mathbb{C}^{N_r M \times 1}$ is the received signal; $\mathbf{H}\in \mathbb{C}^{N_r M \times N_s}$ is the channel matrix; $\mathbf{\Phi}=\operatorname{diag} (\boldsymbol{\phi})\in \mathbb{C}^{N_s \times N_s}$, $\boldsymbol{\phi}=[e^{j\phi_1},\ldots,e^{j\phi_{N_s}}]^{T}\in \mathbb{C}^{N_s\times 1}$ is the phase vector of HMIMOS part composed of $N_s$ patch antennas.  $ {p_u}$ is the transmitted power;  $\mathbf{V}\in \mathbb{C}^{N_s \times N_r M}$ is the precoding matrix; $\mathbf{x} \in \mathbb{C}^{ N_r M \times 1}$ is the transmitted signal, and $\mathbf{w}\in \mathbb{C}^{N_r M \times 1}$ is additive Gaussian noise that has i.i.d. elements with zero mean and variance $\sigma_w^2$. 

Let $\mathbf{y}_a= \mathbf{U}_r^{H} \mathbf{y}\in \mathbb{C}^{n_r M \times 1}$ the received signal in wavenumber domain, $\tilde{\mathbf{H}}_a=\mathbf{H}_a \mathbf{U}_s^{H} \mathbf{\Phi} \in \mathbb{C}^{ n_r M\times N_s}$ is the equivalent channel that incorporates the phase matrix. Thus, 
\begin{equation}
		\mathbf{y}_a =\sqrt{p_u} \tilde{\mathbf{H}}_a    \mathbf{V} \mathbf{x}+\mathbf{w}.
\end{equation}

The final signal received by $m$-th user at $i$-th received sampling point  can be given by
\begin{equation}
\begin{aligned}
	[{y}^{(m)}_a]_i=&[\tilde{\mathbf{H}}^{(m)}_a]_{i,:}   [\mathbf{V}^{(m)} ]_{:,i}   {x}^{(m)}_{i} \\
	&+  [\tilde{\mathbf{H}}^{(m)}_a]_{i,:}    \sum_{i'\neq i}^{n_r}   [\mathbf{V}^{(m)} ]_{:,i'}  {x}^{(m)}_{i}   +  {w}^{(m)}_{i},
\end{aligned}	
\end{equation}
where $\mathbf{x}^{(m)}=[x^{(m)}_{1} , \ldots, x^{(m)}_{N_r} ]\in \mathbb{C}^{N_r \times 1}$ is the transmitted signal of $m$-th user, with $x^{(m)}_{i}$ the $i$-th entry.  $ \mathbf{V}^{(m)}\in \mathbb{C}^{N_s \times N_r}$ is the $m$-th sub-block in precoding matrix corresponding to the $m$-th user, and $ [\mathbf{V}^{(m)} ]_{:,i} $ is the $i$-th column. $\tilde{\mathbf{H}}^{(m)}_a$ is the $m$-th sub-block of the channel channel matrix $\tilde{\mathbf{H}}_a $, and $[\tilde{\mathbf{H}}^{(m)}_a]_{i,:} $ is the $i$-th row.  $\mathbf{y}^{(m)} \in \mathbb{C}^{N_r \times 1}$ is the received signal of $m$-th user, and $[{y}^{(m)}_a]_i$ is the $i$-th element.  

ZF precoding scheme intends at eliminating interference among different users by setting the precoding matrix as $\mathbf{V}=\alpha_{\rm ZF} \tilde{\mathbf{H}}_a^{H} \left(\tilde{\mathbf{H}}_a  \tilde{\mathbf{H}}_a^{H}\right)^{-1} $, where $\alpha_{\rm ZF}$ is the normalization factor to obey  the constraint $\mathbb{E}\{ \operatorname{Tr} ( \mathbf{V}    \mathbf{V} ^{H}  )\}=1$. In this case, it holds $\alpha_{\rm ZF} \tilde{\mathbf{H}}^{(m)}_a \mathbf{V}^{(m)}=\mathbf{I}_{n_r}$. Specifically, we adopt the vector normalization method in \cite{6477575}, i.e., $\tilde{\mathbf{H}}_a^{H} \left(\tilde{\mathbf{H}}_a  \tilde{\mathbf{H}}_a^{H}\right)^{-1}=[\mathbf{f}_1, \ldots, \mathbf{f}_{n_r} ]$,  $\mathbf{V}_{:,i}=\frac{ \mathbf{f}_i}{\sqrt{n_r} \| \mathbf{f}_i \| }$. Thus, the achievable rate is given by 
\begin{equation}  
	\begin{aligned}
		&\mathcal{R} ^{(m),i}_{({\rm ZF})} \! =\!\mathbb{E} \left \{ \log_2\left(1\! +\! \frac{p_u   |[  \mathbf{H}^{(m)}_a]_{i,:} [ \mathbf{H}^{(m)}_a]^{-1}_{:,i}|^2  }{p_u \alpha_{\rm ZF}^2 | [ \mathbf{H}^{(m)}_a]_{i,:}  \sum_{i'\neq i}^{n_r}  [ \mathbf{H}^{(m)}_a]^{-1}_{:,i'}  |^2 \!  + \! \sigma^2_w}\! \right)  \! \right\} \! \\
		&=\mathbb{E} \left \{ \log_2\left(1+\frac{p_u  |[  \mathbf{H}^{(m)}_a]_{i,:} \frac{ \mathbf{f}_{i}}{\sqrt{n_r} \| \mathbf{f}_{i} \| }  |^2  }{p_u   | [ \mathbf{H}^{(m)}_a]_{i,:}  \sum_{i'\neq i}^{n_r}  \frac{ \mathbf{f}_{i'}}{\sqrt{n_r} \| \mathbf{f}_{i'} \| }    |^2  + \sigma^2_w}\right)  \right\} \\
		&= \mathbb{E} \left \{ \log_2\left(1+ {  \frac{p_u}{ {n_r} \sigma^2_w \| \mathbf{f}_i \|^2 }       } \right)  \right\}.  
	\end{aligned}
\end{equation}

We adopt the mathematical method in \cite{wong2008array} to analyze the theoretical capacity. Due to $\mathbf{f}_i=\tilde{\mathbf{H}}_a^{H} \left(\tilde{\mathbf{H}}_a  \tilde{\mathbf{H}}_a^{H}\right)^{-1} \mathbf{e}_{i}$, where $\mathbf{e}_{i}$ is a column vector that is $1$ at $i$-th entry and $0$ otherwise, the term $\frac{1}{   \| \mathbf{f}_i \|^2 }$ can be given as 

\begin{equation}
	\begin{aligned}
		\beta_{i}&=\frac{1}{   \| \mathbf{f}_i \|^2 } =\frac{1}{   \| \tilde{\mathbf{H}}_a^{H} \left(\tilde{\mathbf{H}}_a  \tilde{\mathbf{H}}_a^{H}\right)^{-1} \mathbf{e}_{i} \|^2 }\\
		&= \frac{1}{ \mathbf{e}_{i}^{T} \left( {\mathbf{H}}_a   {\mathbf{H}}_a^{H}\right)^{-1} \mathbf{e}_{i} } =\frac{\operatorname{det} [\mathbf{H}_a \mathbf{H}_a^{H}] }{\operatorname{det} [\mathbf{H}_a^{(i)-} [\mathbf{H}_a^{(i)-}]^{H}] },
	\end{aligned}
\end{equation}
where $\mathbf{H}_a^{(i)-}$ is the matrix of $\mathbf{H}_a$ deleting $i$-th row, and $\operatorname{det} $ denotes the determinant of a matrix. We have 
\begin{equation}
	\begin{aligned}
		\operatorname{det} [\mathbf{H}_a \mathbf{H}_a^{H}] = \sum_{i=1}^{n_r} (-1)^{i-1} \mathbf{g}_1^{H} \mathbf{g}_i   \operatorname{det} [\mathbf{H}_a^{(1)-} [\mathbf{H}_a^{(i)-}]^{H}]. 
	\end{aligned}
\end{equation}

Thus,
\begin{equation}
	\begin{aligned}
		&\beta_{1}  =\mathbf{g}_1^{H} \mathbf{g}_1 - \frac{ \sum_{i=2}^{n_r} (-1)^{i} \mathbf{g}_1^{H} \mathbf{g}_i   \operatorname{det} [\mathbf{H}_a^{(1)-} [\mathbf{H}_a^{(i)-}]^{H}]} {\operatorname{det} [\mathbf{H}_a^{(1)-} [\mathbf{H}_a^{(1)-}]^{H}] }\\
		&=\!\mathbf{g}_1^{H}\! \mathbf{g}_1 \! -\! \sum_{i=2}^{n_r}\! (\!-1\!)^{i}\! \mathbf{g}_1^{H}\! \mathbf{g}_i  \frac{\sum_{k=2}^{n_r} (-1)^{k-1} \mathbf{g}_k^{H} \mathbf{g}_1 \operatorname{det} [\mathbf{M}_k]  }{\!\sum_{j=2}^{n_r} (-1)^{i+j-1} \mathbf{g}_j^{H} \mathbf{g}_i \operatorname{det} [\mathbf{M}_j]\! }\\
		&=\mathbf{g}_1^{H} \mathbf{g}_1 -  \sum_{i=2}^{n_r}  \frac{\sum_{k=2}^{n_r} (-1)^{k-1} \mathbf{g}_k^{H} \mathbf{g}_1  \mathbf{g}_1^{H} \mathbf{g}_i  \operatorname{det} [\mathbf{M}_k]  }{\sum_{j=2}^{n_r} (-1)^{j-1} \mathbf{g}_j^{H} \mathbf{g}_i \operatorname{det} [\mathbf{M}_j] },
	\end{aligned}
\end{equation}
where $\mathbf{M}_k$ is the sub-block matrix of $\mathbf{H}_a^{(1)-} [\mathbf{H}_a^{(1)-}]^{H}$ with removal of the $k$-th row and the $i$-th column. Using the fact that each element in $\mathbf{H}_a$ is independent, and $\mathbb{E}\{[\mathbf{H}_a]_{i,j}^2\}=\sigma_{r,i}^2 \sigma_{s,j}^2$ under separable scattering environment, we have  $\mathbb{E}\{[\mathbf{g}_1 \mathbf{g}_1^H]\}=\operatorname{diag} [\sigma_{r,1}^2 \sigma_{s,1}^2,\ldots,\sigma_{r,1}^2 \sigma_{s,n_s}^2]$. Thus,  
\begin{equation}
	\begin{aligned}
		&\mathbb{E} \left\{ \frac{\sum_{k=2}^{n_r} (-1)^{k-1} \mathbf{g}_k^{H} \mathbf{g}_1  \mathbf{g}_1^{H} \mathbf{g}_i  \operatorname{det} [\mathbf{M}_k]  }{\sum_{j=2}^{n_r} (-1)^{j-1} \mathbf{g}_j^{H} \mathbf{g}_i \operatorname{det} [\mathbf{M}_j] } \right\}\\
		&=\mathbb{E} \left\{ \frac{\sum_{k=2}^{n_r} (-1)^{k-1} \mathbf{g}_k^{H}\mathbb{E}\{ \mathbf{g}_1  \mathbf{g}_1^{H} \} \mathbf{g}_i  \operatorname{det} [\mathbf{M}_k]  }{\sum_{j=2}^{n_r} (-1)^{j-1} \mathbf{g}_j^{H} \mathbf{g}_i \operatorname{det} [\mathbf{M}_j] } \right\}\\
		& \overset{(a)}{=} {\sigma}_{r,1}^2 \hat{\sigma}_s^2,  
	\end{aligned}
\end{equation}
where (a) simplifies the derivation  using average variances, i.e., $\mathbb{E}\{[\mathbf{g}_1 \mathbf{g}_1^H]\}= {\sigma}_{r,1}^2 \hat{\sigma}_s^2 \mathbf{I}_{n_s}$. As a result, 
\begin{equation}
	\begin{aligned}
		&\mathbb{E}\{\beta_{1}\}= \mathbb{E}\{\mathbf{g}_1^{H} \mathbf{g}_1\} \\
		&\quad -  \mathbb{E}\left\{\sum_{i=2}^{n_r}   \frac{\sum_{k=2}^{n_r} (-1)^{k-1} \mathbf{g}_k^{H} \mathbf{g}_1  \mathbf{g}_1^{H} \mathbf{g}_i  \operatorname{det} [\mathbf{M}_k]  }{\sum_{j=2}^{n_r} (-1)^{j-1} \mathbf{g}_j^{H} \mathbf{g}_i \operatorname{det} [\mathbf{M}_j] } \right\} \\
		&\approx \sigma_{r,1}^2 \sum_{j=1}^{n_s} \sigma_{s,j}^2 -(n_r-1)  {\sigma}_{r,1}^2 \hat{\sigma}_s^2\\
		&= (n_s-n_r+1) \sigma_{r,1}^2 \hat{\sigma}_{s}^2.  
	\end{aligned}
\end{equation}

Based upon the above observation, we can derive the theoretical capacity as 
\begin{equation}  
	\begin{aligned}
		\tilde{\mathcal{R}} ^{(m),i}_{({\rm ZF})}  \approx \log_2\left(1+ {  \frac{p_u  }{ {n_r} \sigma^2_w  }   (n_s-n_r+1) \sigma_{r,i}^2 \hat{\sigma}_{s}^2       } \right). 
	\end{aligned}
\end{equation}


 \section{Performance Evaluation}\label{sec:simu}
In this section, we present computer simulation results of the downlink SE in the considered MU-HMIMOS system as well as the theoretical capacity. The single-sided correlation \cite{6297472} and three users are considered. All capacity curves were obtained after averaging over $800$ independent Monte Carlo channel realizations.  

\begin{figure} \vspace{-1mm}
	\begin{center}
		\includegraphics[width=0.48\textwidth]{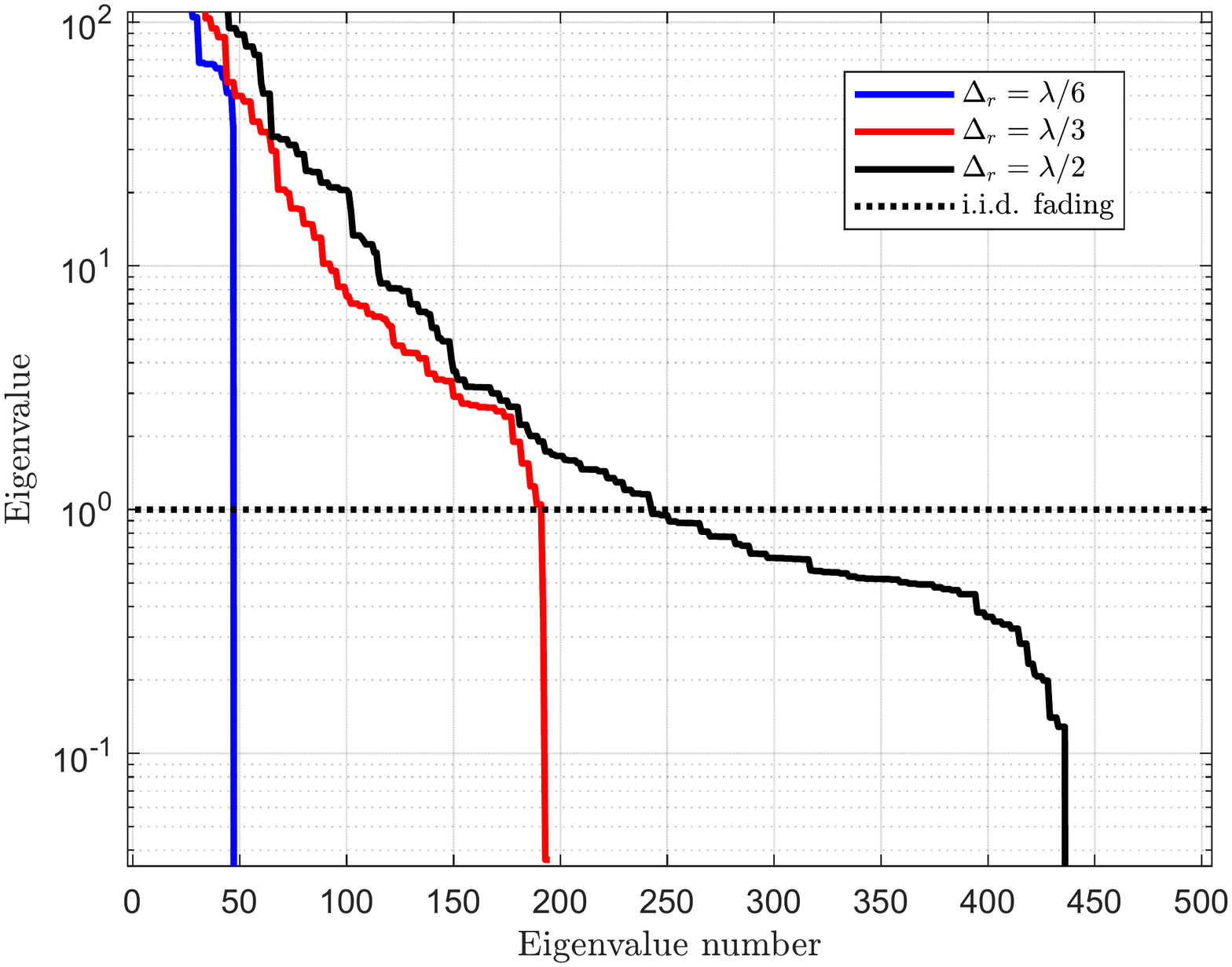}  \vspace{-2mm}
		\caption{The eigenvalues of $\mathbf{R}$ in decreasing order for MU-HMIMOS communication system with $N_s=900,N_r=576, \Delta_s=\lambda/3$ and $\Delta_r\in \{\lambda/6,\lambda/3,\lambda/2\}$.}
		\label{fig:EigvaluesCorNr576} \vspace{-7mm}
	\end{center}
\end{figure}

Fig$.$~\ref{fig:EigvaluesCorNr576} illustrates the eigenvalues of channel correlation matrix $\mathbf{R} $ in decreasing order in a setup with $N_s=900,N_r=576, \Delta_s=\lambda/3$ for different spacing in received patch antennas. From the figure,  eigenvalues are large but non-identical initially, and then the eigenvalues quickly approach zero. This means the strengths of the coupling coefficients are not all equal even in isotropic propagation, which shows that the MU-HMIMOS channel exhibits spatial correlation. In addition, the smaller spacing among patch antennas, the more uneven the coupling coefficients and the steeper the eigenvalues decay, which implies stronger correlation. Specifically, the curve with $\Delta_r=\lambda/2$ decays much slower than that with $\Delta_r=\lambda/6$. The i.i.d. Rayleigh is also showed in black dot curve as reference. We can see from the figure that none of the curves resembles the reference case, even the curve $\Delta_r=\lambda/2$ that is the closest one still has a major difference. These observations all prove that an EM channel in MU-HMIMOS systems should not adopt i.i.d. Rayleigh fading modeling. 

\begin{figure} \vspace{-1mm}
	\begin{center}
		\includegraphics[width=0.48\textwidth]{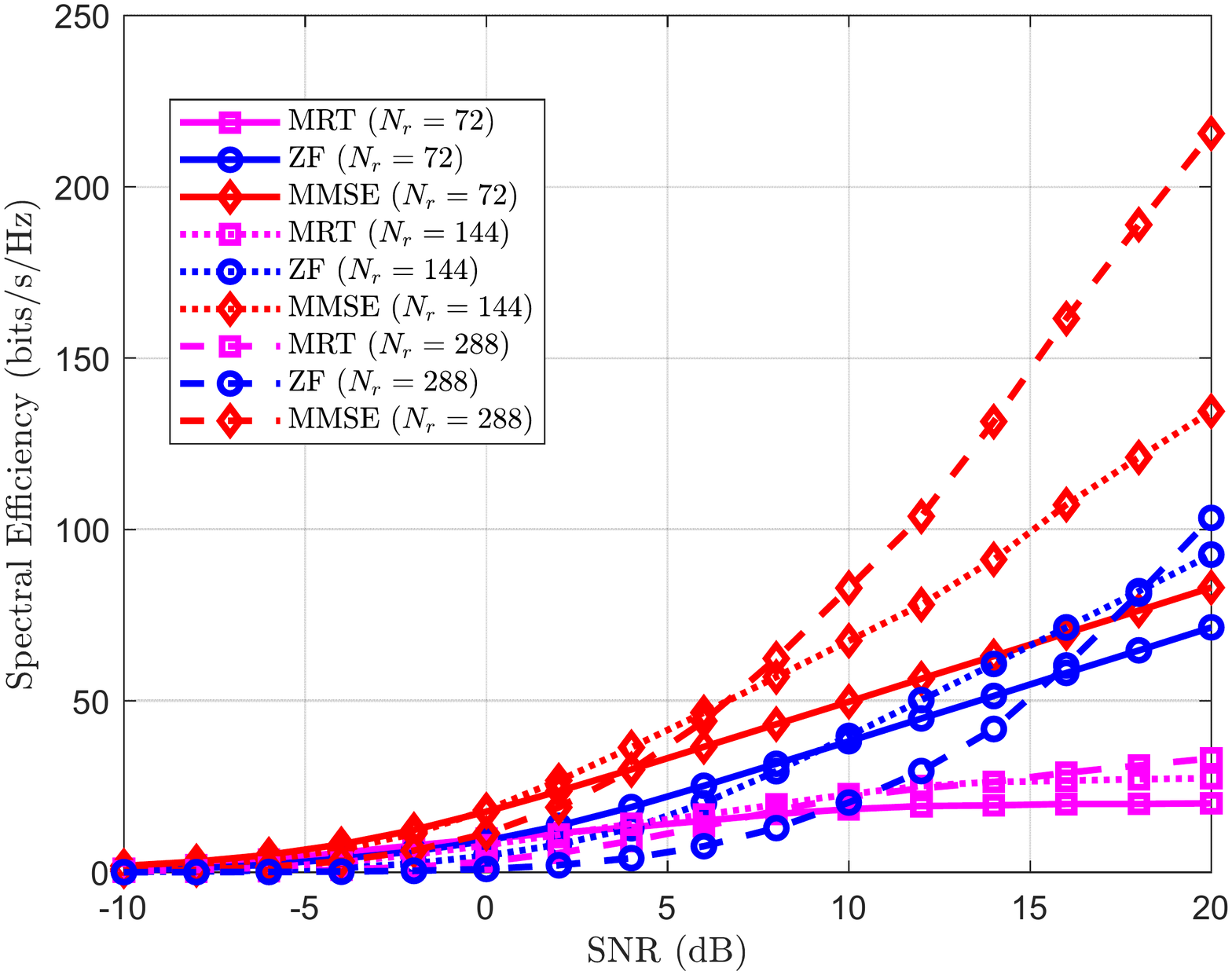}  \vspace{-2mm}
		\caption{SE of MRT, ZF and MMSE precoding schemes for different received patch antennas with $N_s=900,\Delta_s=\Delta_r=\lambda/6$.}
		\label{fig:Ns900s6r6} \vspace{-7mm}
	\end{center}
\end{figure}

\begin{figure} \vspace{-1mm}
	\begin{center}
		\includegraphics[width=0.48\textwidth]{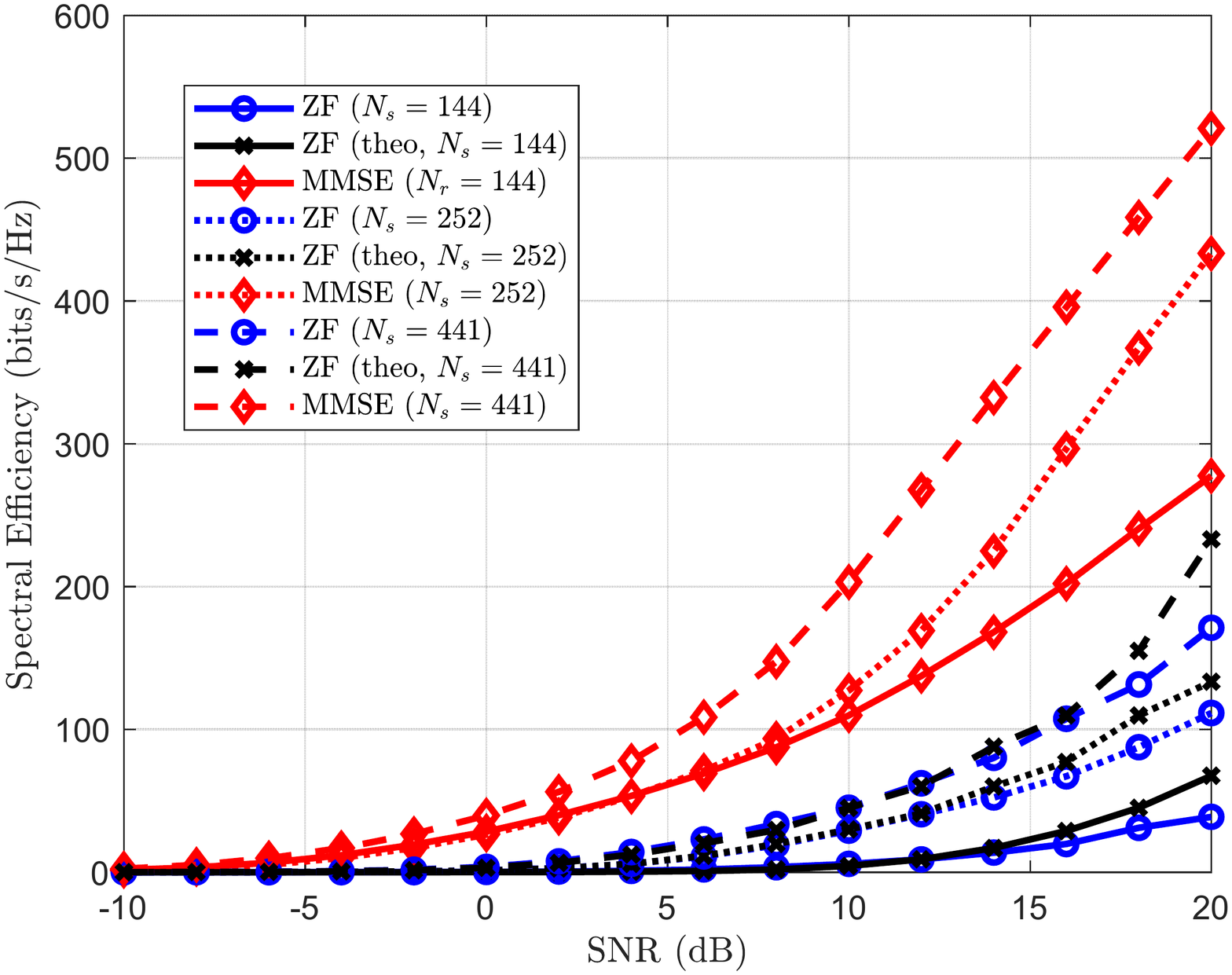}  \vspace{-2mm}
		\caption{SE of ZF precoding, theoretical ZF precoding and MMSE precoding schemes for different transmit patch antennas with $N_r=144, \Delta_s= \Delta_r=\lambda/3$.}
		\label{fig:Nr144s3r3_MMSE_ZF} \vspace{-7mm}
	\end{center}
\end{figure}

The impact of received patch antennas on the SE of MRT, ZF and MMSE precoding schemes for $N_s=900, \Delta_s=\Delta_r=\lambda/6$ is given in Fig.~\ref{fig:Ns900s6r6}.   As shown in figure,  the MRT precoding is better than ZF precoding in the low SNR region, which is contrary to the case in the high SNR region. This is mainly due to the noise, i.e., the noise is dominant in low SNR region, thus, MRT is better, while the noise impact is finite in the high SNR region, thus, ZF is better. As a benchmark, MMSE performs the best in the whole SNR region, and the gap between ZF and MMSE gradually decreases with the increase of SNR.  In addition, the more received patch antennas bring benefits in SE. As observed from figure, the case $N_r=288$ achieves the best performance compared with $N_r=72$ and $N_r=144$. This can be accounted for the larger received surface area enlarged by more received patch antennas under the fixed spacing. 
Fig$.$~\ref{fig:Nr144s3r3_MMSE_ZF} shows the SE of ZF precoding and theoretical ZF precoding schemes with $N_r=144, \Delta_s=\Delta_r=\lambda/3$ for different number of transmit patch antennas, and MMSE precoding scheme is set as benchmark curve. As shown in figure, there is very small gap between the ZF precoding and theoretical ZF schemes. Specifically, at lower SNR, the theoretical ZF perfectly predict the ZF precoding scheme. Thus, the effectiveness of the presented theoretical capacity is proved. Naturally, the more transmit patch antennas bring more benefits in SE. This can be accounted for that transmit surface is larger with the increase of patch antennas given the fixed spacing. What's more, the gap between ZF precoding and  benchmark MMSE precoding is narrower with the increase of transmit patch antennas in high SNR region. This is mainly because the noise has little impact on ZF precoding when SNR goes high.  

\begin{figure} \vspace{-1mm}
	\begin{center}
		\includegraphics[width=0.48\textwidth]{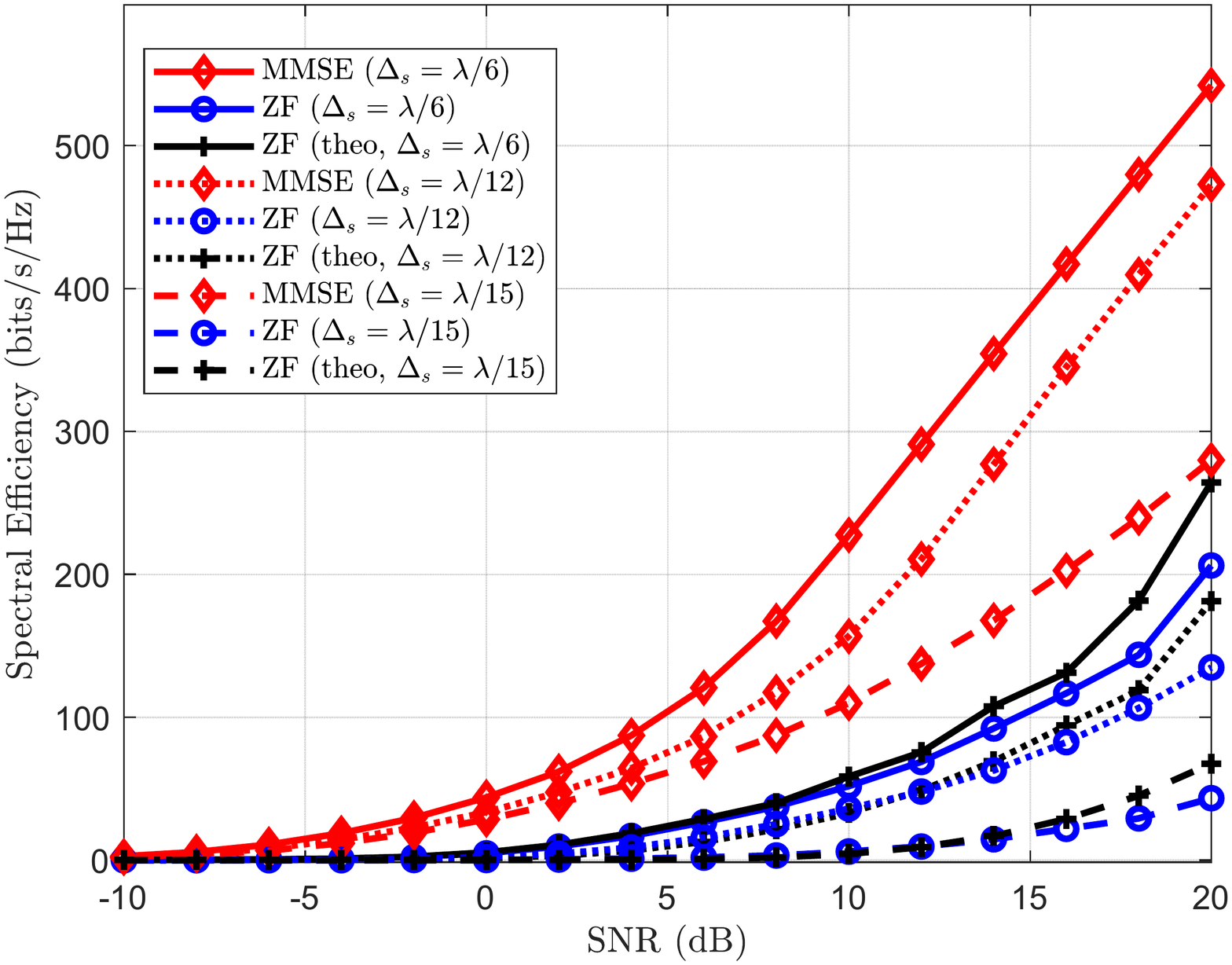}  \vspace{-2mm}
		\caption{SE of ZF and MMSE precoding schemes for different spacing with $N_s=3600,N_r=144,\Delta_r=\lambda/3$.}
		\label{fig:Ns3600Nr144r3_MMSE_ZF} \vspace{-7mm}
	\end{center}
\end{figure}

The impact of transmit spacing on SE for MMSE, ZF, and theoretical ZF  precoding schemes are given in Fig$.$~\ref{fig:Ns3600Nr144r3_MMSE_ZF}, under the settings of  $N_s=3600,N_r=144,\Delta_r=\lambda/3$. From the Fig$.$~\ref{fig:Ns3600Nr144r3_MMSE_ZF}, smaller spacing with giving number of patch antennas has less surface area and induces more correlation among patch antennas, thus, the SE is worse. Specifically, the ZF precoding scheme  with $\Delta =\lambda/15$ has worse performance than  $\Delta =\lambda/6$. In other words, there is a large reduction in correlated channel with larger spacing. Normally, the spacing $\Delta_s=\Delta_r=\lambda/2$ is adopted in the uncorrelated channel assumption for SE analysis \cite{6457363}. In addition, it is shown that the theoretical ZF performance with closer spacing (higher correlation) predicts the ZF precoding sufficiently well. 

\section{Conclusion}\label{sec:conclusion}
In this paper, we presented a simple and convenient channel model for MU-HMIMOS communication systems. The proposed model is expressed in the wavenumber domain, thus, is explicitly describes the dependence of signal propagation on the amount and spacing of the HMIMOS patch antennas. We also have analytically investigated the ZF precoding scheme, presenting a theoretical SE analysis. Our simulation results verified that the more patch antennas with fixed spacing are present at the transmitter and receiver, the more the achievable SE improves when the surface gets larger. However, smaller antenna spacing induces stronger correlations in system, which can impair the SE performance of various precoding schemes. In addition, it was shown that the presented theoretical analysis performs well even in highly correlated scenarios. 

\section*{Acknowledgments}
The work of Prof. Huang was supported by the China National Key R\&D Program under Grant 2021YFA1000502, National Natural Science Foundation of China under Grant 62101492, Zhejiang Provincial Natural Science Foundation of China under Grant LR22F010002, National Natural Science Fund for Excellent Young Scientists Fund Program(Overseas), Ng Teng Fong Charitable Foundation in the form of ZJU-SUTD IDEA Grant, Zhejiang University Education Foundation Qizhen Scholar Foundation, and Fundamental Research Funds for the Central Universities under Grant 2021FZZX001-21.  This research is supported by Singapore Ministry of Education Tier 2 MOE-000168-01.   The work of Prof. G. C. Alexandropoulos was supported by the EU H2020 RISE-6G project under grant number 101017011.

\bibliographystyle{IEEEbib}
\bibliography{strings}
\end{document}